\documentclass[preprint,aps]{revtex4}
 \usepackage{amssymb} \usepackage{graphicx} \graphicspath{ {images/} }
  \usepackage{dcolumn} \usepackage{amsmath}
\usepackage{tikz}
  \usetikzlibrary{shapes,backgrounds}
\usepackage{color}

\begin{document}

\title{Motion of membranes in space-times with torsion}
\author{\"{O}zg\"{u}r A\c{c}{\i}k $^{1}$}
 \email{ozacik@science.ankara.edu.tr}
\author{Aytolun \c{C}atalkaya $^{1}$}
\email{aytolun.catalkaya@hotmail.com}
\author{\"{U}mit Ertem $^{2}$} 
\email{umertemm@gmail.com, umit.ertem@tedu.edu.tr}
\author{\"{O}zg\"{u}n S\"{u}temen $^{1}$}
\email{ozgun.sutemen@gmail.com}

\address{$^{1}$ Department of Physics,
Ankara University, Faculty of Sciences, 06100, Tando\u gan-Ankara,
Turkey\\
$^{2}$ TED University, Ziya G\"{o}kalp Caddesi, No:48, 06420, Kolej \c{C}ankaya, Ankara,
Turkey}

\begin{abstract}
The motion of membranes interacting with external fields in space-times with curvature and torsion is considered. The intrinsic and extrinsic properties of the immersion are fused together to form a stress tensor for the corresponding material hypersurface. This geometro-elastic stress tensor is part of the total stress tensor by which it looses the symmetry and divergenceless properties because of the existence of torsion. The equation of motion of the membrane is given by equating the total stress tensor to a non-zero value determined by the curvature and torsion of the ambient space-time. Dirac and \"Onder-Tucker bubbles are considered as special cases. An example of the membrane motion on a manifold admitting a generalized Killing spinor is given.

\end{abstract}

\maketitle

\section{Introduction}
In 1962, Dirac considered an extended model (bubble model) for the electron, the first excited state of which was described as the muon corresponding to the changes of the shape and size of the bubble \cite{Dirac}. Preceding Dirac, the shape and size fixed, extended general relativistic electron was considered by Lees \cite{Lees}, which contextually by itself rules out the excited states of the electron. Dirac had thought that the surface tension of the electron in its rest frame could balance the electrical repulsion; but later on it was found that the Dirac bubble was not stable against quadruple deformations \cite{Gnadig et al}. In another paper, \"{O}nder and Tucker generalized Dirac's work by a model that can be put into the perspective of a class of Willmore-type immersions \cite{Onder Tucker}. Their contribution came from adding a term to Dirac's action including the extrinsic curvature of the immersion. Extrinsic curvature terms appear in most theories of relativistic extended objects described by effective Lagrangian actions \cite{Gregory}. In \cite{Tucker mem} the equation of a membrane interacting with external fields in space-time with curvature was established. Ref. \cite{Tucker mem} generalizes Dirac's work in geometrical terms as follows: Dirac encoded the role of surface tension by defining a stress tensor proportional to the induced metric on the hypersurface, though in \cite{Tucker mem} the additional geometrical term included the shape tensor of the immersion so that also the external geometrical properties of the membrane become significant. It is important to note that the generality of the form for the stress tensor of the immersion in this model reduces to the model given in \cite{Onder Tucker} for a special choice of geometric invariants of the membrane history. A crucial analysis both for the classical and quantum motion of free relativistic membranes was done in \cite{Collins Tucker}. For the quantum regime, they developed a covariant formalism with constraints and an alternative theory in terms of independent degrees of freedom. Later on \"{O}nder and Tucker investigated the semiclassical dynamics of a charged relativistic membrane resulting with a harmonic estimate of radial quantum modes \cite{Onder Tucker 2}; but their work was refined and improved in \cite{Cordero Molgado Rojas} by using a classical canonical approach relying on the Ostrogradski-Hamiltonian formalism and then applying the Dirac's constraint quantization scheme.

In this work we aim to generalize \cite{Tucker mem} by including the space-time torsion and deduce the possible contributions on the equation of motion of a relativistic membrane subject to external fields in background spacetimes. Here the main scheme is based on a distribution valued total stress energy momentum tensor, by which it loses its symmetry and divergenceless properties because of the presence of torsion. So the equation of motion for the membrane is determined by equating the divergence of the stress tensor to a non-zero value, that is constructed out of curvature and torsion of the ambient space-time. Since \cite{Tucker mem} includes the results of \cite{Dirac} and \cite{Onder Tucker} as special cases, we also investigate the motions of both membranes which we call Dirac bubble and \"{O}nder-Tucker bubble respectively. The extra terms coming from the existence of torsion are determined for both cases. We also examplify our investigations by considering a specific kind of space-time which admits a generalized Killing spinor field and a related non-zero torsion. Then, it is seen that the corresponding equations of motion simplify in terms of the Killing function; so it is easy to see the limiting case for vanishing torsion by taking this function to zero.

The organization of the paper is as follows. In section II we introduce the distribution valued total stress tensor and give the necessary differential operations in the language of coordinate-free differential geometry. After defining the equation of motion of the membrane by equating the divergence of the total stress tensor to a non-zero value determined by the curvature and torsion of the background, then  we deduce the normal and tangential jump conditions for the local physical momentum $1$-form current. These conditions are given in terms of the divergence of the hypersurface stress tensor. The (immersed) hypersurface is a time-like $3$-manifold corresponding to the space-time history of the membrane. In section III, a convenient characterisation of the hypersurface stress tensor is given out of the extrinsic and intrinsic geometric properties of the immersion and accompanying invariant functions. Here the topology of the membrane is restricted to be spherical and this bubble is coupled to an external electromagnetic field. The reduction to the Dirac bubble case is analyzed in section III A and to the \"{O}nder-Tucker bubble case in section III B. Section IV is devoted to the examplification of the previous results in a space-time admitting a generalized Killing spinor. The conclusion of our results constitutes section V. There are two following appendices that we feel necessary for the completeness of the paper; where Appendix A contains some knowledge about curved space-time distributions and Appendix B gives the Gauss-Codazzi formalism for codimension $1$ immerions in the presence of torsion.

\section{Stress-energy-momentum tensors with discontinuities}
Let $C:D\rightarrow M$ is a time-like 3-dimensional immersion into a 4-dimensional non-Riemannian space-time $(M,g,\nabla)$, where $D$ is some 3-dimensional parameter domain, $g$ is the space-time metric with Lorentzian signature and $\nabla$ the metric compatible connection with torsion.
The induced metric and the induced connection on the image $\Sigma$ of $C$ are denoted respectively as $\widehat{g}$ and $\widehat{\nabla}$. If the local equation of the hypersurface is given by $\Phi=0$ then the other parts of space-time are labelled by $+$ and $-$ corresponding to $\Phi>0$ and $\Phi<0$; so $M=M^{+}\cup M^{-} \cup M^{0}$ where $M^{0}=\Sigma=\partial M^{-}=-\partial M^{+}$ for convenient boundary conditions at spatial infinity. The unit space-like normal $N$ of the regular hypersurface $\Sigma$ is metrically related to the 1-form $d\Phi/|d\Phi|$ where $| |$ denotes the norm with respect to $g$. Usually a tilda over a tensor will denote the tensor associated to it by the space-time metric tensor $g$ so that $\widetilde{N}=d\Phi/|d\Phi|$. For rank 2 covariant tensors we also have other usages for tilda where in a chart $x=(x^{\mu})$ if $G=G_{\mu\nu}dx^{\mu}\otimes dx^{\nu}$ then we define the associated contravariant rank 2 tensor as $\widetilde{G}=G_{\mu\nu}\widetilde{dx^{\mu}}\otimes \widetilde{dx^{\nu}}$ and the associated mixed rank 2 tensors as $^{\widetilde{}}G=G_{\mu\nu}\widetilde{dx^{\mu}}\otimes dx^{\nu}$ and $G ^{\widetilde{}}=G_{\mu\nu}dx^{\mu}\otimes \widetilde{dx^{\nu}}$. The Heaviside function with support on $\Phi > 0$ is $\Theta^{+}$ and the scalar Dirac distribution $\delta(\Phi)$ is given by the following relation
\begin{eqnarray}
d\Theta^{+}=\delta(\Phi) d\Phi
\end{eqnarray}
and we suppose that the g-dual of the total stress-energy-momentum tensor takes the form
\begin{eqnarray}
\widetilde{T}=\widetilde{T}_{+}\Theta^{+}+\widetilde{T}_{-}\Theta^{-}+\widetilde{T}_{0}\,\delta(\Phi)
\end{eqnarray}
such that the coefficient tensors are all smooth \cite{Lichnerowicz2}. Here total stress-energy-momentum tensor $T$ contains all matter and fields and $T_{0}$ describes the stress properties of the material hypersurface $\Sigma$.

The covariant divergence of a bidegree (2,0) tensor (i.e. with contravariant rank 2 and covariant rank 0) $\widetilde{T}$ is a (1,0) tensor $\nabla.\widetilde{T}$ and can be defined as
\begin{eqnarray}
\nabla.\widetilde{T}=(\nabla_{X_a}\widetilde{T})(e^a,-)
\end{eqnarray}
where $\{X_a\}$ and $\{e^a\}$ being arbitrary dual basis. If $f$ is a scalar on $M$ then
\begin{eqnarray}
\nabla.(f\widetilde{T})&=&e^a(\nabla.(f\widetilde{T}))X_a=(\nabla_{X_b}(f\widetilde{T})(e^b,e^a))X_a\\
&=&\widetilde{T}(df,e^a)X_a+f((\nabla_{X_b}\widetilde{T})(e^b,e^a))X_a \nonumber
\end{eqnarray}
so we get
\begin{eqnarray}
\nabla.(f\widetilde{T})=\widetilde{T}(df,-)+f\nabla.\widetilde{T}.
\end{eqnarray}
Because Einstein tensor $G$ has non-zero covariant divergence in a dynamical Riemann-Cartan space-time, that is given by
\begin{eqnarray}
\nabla.G=-*^{-1}(T^q\wedge i_{X_q}R_{ab}\wedge *{e^{ab}}_{p})e^p.
\end{eqnarray}
Here $*$ is the Hodge map associated with $g$, $T^a$'s are torsion 2-forms and ${R^a}_b$'s are the curvature 2-forms of $\nabla$. The equation of motion of a hypersurface with possible (non-gravitational) interactions through the regions $+$ and $-$ may be given as
\begin{eqnarray}
\nabla.\widetilde{T}=-*^{-1}(T^q\wedge i_{X_q}R_{ab}\wedge *{e^{abp}})X_p:=\widetilde{\beta}
\end{eqnarray}
in a background Riemann-Cartan spacetime. It is convenient at this point to say that the torsion may have a spinorial origin such that the spinor field has a vanishing stress tensor; so that it does not contribute to the total stress tensor. Thus if $\nabla.\widetilde{T}_+=0$ in $M^+$ and $\nabla.\widetilde{T}_{-}=0$ in $M^{-}$ we get
\begin{eqnarray}
[\widetilde{T}](d\Phi,-)-\widetilde{\beta}=-(\nabla.\widetilde{T}_{0})|_{\Sigma}
\end{eqnarray}
where naturally
\begin{eqnarray}
\widetilde{T}_{0}(d\Phi,-)|_{\Sigma}=0
\end{eqnarray}
and $[\widetilde{T}]:=\widetilde{T}_+-\widetilde{T}_{-}$ is the discontinuity of $\widetilde{T}$ across the hypersurface. Equation (9) comes from the assumption that $T_0$ is only defined on the hypersurface. Since $d\Phi$ is space-like, equation (8) can be interpreted as the jump in the local (physical) momentum current $[J_{d\Phi}]=-[T(\widetilde{d\Phi},-)]$ which gives the normal force on the hypersurface. If $c$ is a time-like future pointing observer curve then $i_{\dot{c}}*J$ is the local force 2-form belonging to any local momentum current 1-form $J$.\\

 For a general mixed tensor $S$ we may define $\breve{\Pi}_{N}S$ as
\begin{eqnarray}
(\breve{\Pi}_{N}S)(Y_1,Y_2,...,\beta_1,\beta_2,...)=S(\Pi_{N}Y_1,\Pi_{N}Y_2,...,\widetilde{\Pi}_{N}\beta_1,\widetilde{\Pi}_{N}\beta_2,...)
\end{eqnarray}
where $\Pi_{N}=1-\{\widetilde{N}(N)\}^{-1}N\otimes\widetilde{N}$ and $\widetilde{\Pi}_{N}=1-\{\widetilde{N}(N)\}^{-1}\widetilde{N}\otimes N$
are the (1,1) projection tensors and one should remember that $\widetilde{N}(N)=g(N,N)=1$, here $Y_i$'s are vector fields and $\beta_j$'s are $1$-forms in spacetime. $\breve{\Pi}_{N}S$ will be termed as $g$-orthogonal to $N$ or as the hypersurface tensor corresponding to $S$. So with this definition equation (9) says that the hypersurface stress-energy-momentum tensor must be $g$-orthogonal to its normal vector field, i.e.
\begin{eqnarray}
\breve{\Pi}_{N}\widetilde{T}_{0}=\widetilde{T}_{0} .
\end{eqnarray}
The $N$-decomposition of $\nabla.\widetilde{T}_{0}$ is
\begin{eqnarray}
\nabla.\widetilde{T}_{0}=\Pi_{N}(\nabla.\widetilde{T}_{0})+\widetilde{N}(\nabla.\widetilde{T}_{0})N
\end{eqnarray}
where the first term at the right hand side is $g$-orthogonal to $N$ and the second term is parallel to $N$.
We can choose $\{X_a\}$ and its dual $\{e^a\}$ as $X_0$ time-like and $X_1=N$ so from (3) we can write $\widetilde{N}(\nabla.\widetilde{T}_{0})=(\nabla_{X_a}\widetilde{T}_{0})(e^a,\widetilde{N})$ and from
(11) we deduce
\begin{eqnarray}
(\nabla_{X_a}\widetilde{T}_{0})(e^a,\widetilde{N})&=&\nabla_{X_a}(\widetilde{T}_{0}(e^a,\widetilde{N}))-\widetilde{T}_{0}(\nabla_{X_a}e^a,\widetilde{N})-\widetilde{T}_{0}(e^a,\nabla_{X_a}\widetilde{N}) \\
&=&-\widetilde{T}_{0}(e^a,\nabla_{X_a}\widetilde{N}) =-\widetilde{T}_{0}(e^i,\nabla_{X_i}\widetilde{N})  \nonumber
\end{eqnarray}
here $i=0,2,3$ since $\Pi_{N}\{X_a\}=\{X_i\}$. From (B3) it is seen that $\nabla_{X_i}N=-A_{N}X_i$ and also using metric compatibility $\nabla g=0$ we reach
\begin{eqnarray}
\widetilde{N}(\nabla.\widetilde{T}_{0})=\widetilde{T}_{0}(e^i,\widetilde{A_{N}X_i}),
\end{eqnarray}
and by further manipulation we write $\widetilde{T}_{0}(e^i,\widetilde{A_{N}X_i})=T_{0}(X^i,A_{N}X_i)$. If we also $N$-decompose the local physical momentum 1-form $[J_{d\Phi}]$ as
\begin{eqnarray}
[J_{d\Phi}]=\widetilde{\Pi}_{N}([J_{d\Phi}])+([J_{d\Phi}](N))\widetilde{N}
\end{eqnarray}
then by using (8) we find the final fundamental equations of motion for our first investigation: the normal (scalar) part is
\begin{eqnarray}
([J_{d\Phi}]+\beta)(N)=T_{0}(X^i,A_{N}X_i)|_{\Sigma}
\end{eqnarray}
and the tangential (3-vector) part is
\begin{eqnarray}
\Pi_{N}([\widetilde{J_{d\Phi}}]+\widetilde{\beta})=\Pi_{N}(\nabla.\widetilde{T}_{0})|_{\Sigma}.
\end{eqnarray}
 \section{Geometric Elasticity}
A convenient way to assign a tensor $T_{0}$ that satisfies the criterion (11) can be constructed from the first and second fundamental forms of the hypersurface as
\begin{eqnarray}
T_{0}=|d\Phi|\{L_1 \breve{\Pi}_{N}g+L_2 \breve{\Pi}_{N}H\}
\end{eqnarray}
where the scalars $L_i$ are selected as functions of $\kappa_1,\kappa_2,\kappa_3, \Delta_{\widehat{g}}\kappa_1,\Delta_{\widehat{g}}\kappa_2,\Delta_{\widehat{g}}\kappa_3$ such that $\Delta_{\widehat{g}}$ is the hypersurface Laplacian. The invariant quantities $\kappa_j$ are being taken as the elementary symmetric functions of the Weingarten map eigenvalues. If $\{X_i|i=0,2,3\}$ is a local $g$-orthonormal basis of tangent vector fields on $\Sigma$ satisfying $A_{N}X_i=\lambda_i X_i$ then $\kappa_1=\lambda_0+\lambda_2+\lambda_3$, $\kappa_2=\lambda_0\lambda_2+\lambda_2\lambda_3+\lambda_3\lambda_0$ and $\kappa_3=\lambda_0\lambda_2\lambda_3$ \cite{Tucker mem}. Because of the presence of torsion, $T_{0}$ will be non-symmetric with respect to its arguments whenever $L_2$ is non-vanishing. From now on we take the topology $S^2\times\mathbb{R}$ for the parameter domain $D$ and couple the membrane to the Maxwell field $F$ by taking $F_{-}=0$ and $T_+=T_{Maxwell}=T_{ab}e^a \otimes e^{b}$ where $T_{ab}=-\frac{1}{4}g_{ab}{F_{+}}^{cd}{F_{+}}_{cd}-{F_{+}}_{ac}{{F_{+}}^{c}}_{b}$ , then
\begin{eqnarray}
[J_{d\Phi}]=-\frac{1}{2}|d\Phi|*(i_{N}F_{+}\wedge *F_{+}-i_{N}*F_{+}\wedge F_{+})
\end{eqnarray}
where $F_{+}$ is the external Coulomb field of the spherical charged membrane in its proper frame. The proper frame is assumed to be determined by the center of mass of the bubble which is not obviously defined in relativistic space-times either flat or curved \cite{Giulini},\cite{Dixon},\cite{Ehlers and Rudolf}. With this construction $M_{-}$ appears as the world tube \cite{Gray} traced out by the interior region of a space-like bubble. Now we can choose specific forms for $T_{0}$ and investigate the corresponding equations of motion.
\subsection{Dirac Bubble}
For some coupling constant $\kappa$ we take the hypersurface stress tensor as
\begin{eqnarray}
T_{0}&=&\kappa |d\Phi| \breve{\Pi}_{N}g \\
&=&\kappa |d\Phi| (g-\widetilde{N} \otimes \widetilde{N}) \nonumber
\end{eqnarray}
and calculate
\begin{eqnarray}
(\nabla.T_{0})(N)=\kappa \{g(\widetilde{d|d\Phi|},N)-\widetilde{N}(\widetilde{d|d\Phi|})\widetilde{N}(N)\}- \kappa |d\Phi| \nabla.(\widetilde{N} \otimes \widetilde{N})(N) \nonumber
\end{eqnarray}
where the terms in the curly parenthesis annihilate each other. If we use the definition of the divergence for the remaining term we get
\begin{eqnarray}
(\nabla.T_{0})(N)=-\kappa |d\Phi| \nabla_{X_a}(\widetilde{N}\otimes \widetilde{N})(X^a,N) \nonumber
                 =-\kappa |d\Phi| \{\widetilde{\nabla_{X_a}N}(X^a)+g(N,X^a)\widetilde{\nabla_{X_a}N}(N)\}
\end{eqnarray}
and further manipulation gives
\begin{eqnarray}
(\nabla.T_{0})(N)=-\kappa |d\Phi| \{g(\nabla_{X_a}N,X^a)+g(\nabla_{N}N,N)\}. \nonumber
\end{eqnarray}
Since $g(\nabla_{N}N,N)=0$ and $g(\nabla_{X_a}N,X^a)=-g(N, \nabla_{X_a}X^a)=-g(N, \nabla_{\Pi_{N}X_a}\Pi_{N}X^a)$ we get
\begin{eqnarray}
(\nabla.T_{0})(N)=\kappa |d\Phi| g(h(\Pi_{N}X_a,\Pi_{N}X^a),N)=\kappa |d\Phi| H(\Pi_{N}X_a,\Pi_{N}X^a).
\end{eqnarray}
Remembering that the mean curvature normal of the hypersurface is $\eta:=\frac{1}{3}h(\Pi_{N}X_a,\Pi_{N}X^a)=\frac{1}{3}\mathcal{H}N$ and with a little algebra
\begin{eqnarray}
[J_{d\Phi}](N)=-\frac{1}{2}|d\Phi|*(-i_{N}F_{+}\wedge *i_{N}F_{+}+F_{+}\wedge \widetilde{N}\wedge*(F_{+}\wedge \widetilde{N})).
\end{eqnarray}
Now we can use the formula for any $p$-forms $\alpha$ and $\beta$ and vector fields $X$ and $Y$
$$i_{X} \alpha \wedge *i_{Y} \beta=g(X,Y)g_p(\alpha,\beta)*1-(\alpha \wedge \widetilde{Y})\wedge *(\beta \wedge \widetilde{X})$$
for further manipulation and use the fact that $F_{+}\wedge \widetilde{N}$ is vanishing for spherical topology in bubble's (local) rest frame. Also using $**1=-1$ for Lorentzian signature in four dimensions and $g_2(F_{+},F_{+}):=*^{-1}(F_{+}\wedge *F_{+})$, then the equation of motion for our membrane is given by
\begin{eqnarray}
 \kappa \mathcal{H}+*^{-1}(T^a\wedge i_{X_a}R_{ij}\wedge * {e^{ij}}_{1})/|d\Phi|=*^{-1}(F_{+}\wedge *F_{+})/2.
\end{eqnarray}
This equation is the generalization of Dirac's \cite{Dirac} to the case for the existence of torsion.
\subsection{\"{O}nder-Tucker Bubble}
This time we set the hypersurface stress tensor to be
\begin{eqnarray}
T_{0}=\kappa |d\Phi| \{Tr(A_N)\breve{\Pi}_{N}g-\breve{\Pi}_{N}H\}
\end{eqnarray}
which also includes the second fundamental form of the immersion, well this is not a symmetric term in the presence of torsion.
Evaluation of the stress tensor on two vector fields $X$ and $A_NY$ gives
\begin{eqnarray}
T_{0}(\Pi_NX,A_N\Pi_NY)&=&\kappa |d\Phi| \{Tr(A_N)g(\Pi_NX,A_N\Pi_NY)-H(\Pi_NX,A_N\Pi_NY)\} \\
&=&\kappa |d\Phi| \{Tr(A_N)g(A_N\Pi_NY,\Pi_NX)-g(A_N\Pi_NX,A_N\Pi_NY)\}  \nonumber
\end{eqnarray}
Let us contract the first and the last arguments of the Riemann tensor,
\begin{eqnarray}
Ric(\Pi_NY,\Pi_NX)&=&R(X_a,\Pi_N Y,\Pi_N X,X^a)  \\
&=&R(\Pi_NX_a,\Pi_N Y,\Pi_N X,\Pi_NX^a)+R(N,\Pi_N Y,\Pi_N X,N) \nonumber
\end{eqnarray}
from Gauss equation (B4), the first term at the right hand side can be decomposed as $$\widehat{Ric}(\Pi_N Y,\Pi_N X)+g(h(\Pi_NX^a,\Pi_NX),h(\Pi_NY,\Pi_NX_a))-g(h(\Pi_NY,\Pi_NX),h(\Pi_NX^a,\Pi_NX_a)).$$
Let us analyze some terms separately. First recall that,  $$g(h(\Pi_NY,\Pi_NX),h(\Pi_NX^a,\Pi_NX_a))=H(\Pi_NX^a,\Pi_NX_a)g(h(\Pi_NY,\Pi_NX),N)$$ since $Tr(A_N)=\mathcal{H}$ , it can be rewritten as $Tr(A_N)g(A_N\Pi_NY,\Pi_NX)$ this is of course equal to
$Tr(A_N)\breve{\Pi}_Ng(A_N Y,X)$ so $$g(h(\Pi_NY,\Pi_NX),h(\Pi_NX^a,\Pi_NX_a))=Tr(A_N)\breve{\Pi}_Ng(A_N Y,X).$$ We can also expand $A_N \Pi_NX$ as $g(A_N \Pi_NX,\Pi_NX^a)\Pi_NX_a=g(h(\Pi_NX,\Pi_NX^a),N)\Pi_NX_a$ and similarly $A_N \Pi_NY=g(h(\Pi_NY,\Pi_NX^b),N)\Pi_NX_b$ so we get $$g(A_N \Pi_NX,A_N\Pi_NY)=g(h(\Pi_NX,\Pi_NX^a),h(\Pi_NY,\Pi_NX_a)).$$
From (B6) the last equality can be expanded further as
$$g(A_N \Pi_NX,A_N\Pi_NY)=g(h(\Pi_NX^a,\Pi_NX),h(\Pi_NY,\Pi_NX_a))+g(\mathbf{T}(\Pi_NX,\Pi_NX^a),h(\Pi_NY,\Pi_NX_a)).$$
As a result we find that
\begin{eqnarray}
(TrA_N)\check{\Pi}_Ng(A_NY,X)&-&\check{\Pi}_Ng(A_NX,A_NY)=\widehat{Ric}(\Pi_N Y,\Pi_N X)-Ric(\Pi_NY,\Pi_NX) \\ \nonumber &+&g(\mathbf{R}(N,\Pi_NY)\Pi_NX,N)+g(\mathbf{T}(\Pi_NX_a,\Pi_NX),h(\Pi_NY,\Pi_NX^a)) \\   \nonumber
\end{eqnarray}
where the left hand side is easily seen to be $T_{0}(A_NY,X)/(\kappa |d\Phi|)$ hence we get
\begin{eqnarray}
(\nabla.T_{0})(N)&=&\kappa |d\Phi|\{\widehat{\mathcal{R}}-\mathcal{R}+ Ric(N,N)+g(\mathbf{R}(N,\Pi_NX^b)\Pi_NX_b,N)\\   \nonumber
&+&g(\breve{\Pi}_N\mathbf{T}(X_a,X_b),\breve{\Pi}_Nh(X^b,X^a))\}.      \nonumber
\end{eqnarray}
Since $g(\Pi_NX_b,N)=0$ then $\mathbf{R}(N,\Pi_NX^b)g(\Pi_NX_b,N)=0$ which implies that $g(\mathbf{R}(N,\Pi_NX^b)\Pi_NX_b,N)=-g(\Pi_NX_b,\mathbf{R}(N,\Pi_NX^b)N)=\widehat{Ric}(N,N).$
So, the equation of motion of the membrane coupled to Maxwell field in the presence of torsion is
\begin{eqnarray}
\kappa \{\widehat{\mathcal{R}}-\mathcal{R}&+&2 Ric(N,N)+g(\breve{\Pi}_N\mathbf{T}(X_a,X_b),\breve{\Pi}_Nh(X^b,X^a))\} \\
&+&*^{-1}(T^a\wedge i_{X_a}R_{ij}\wedge * {e^{ij}}_{1})/|d\Phi|=*^{-1}(F_{+}\wedge *F_{+})/2. \nonumber
\end{eqnarray}
This equation is the generalization of the equation of motion found by \"{O}nder and Tucker \cite{Onder Tucker} for the presence of torsion.

 \section{The investigation of membrane motions in a specific non-Riemannian spacetime}
We assume that our ambient spacetime admits a general kind of Killing spinor field $\psi \in \Gamma S^{\mathbb{C}}M$ satisfying
\begin{eqnarray}
\nabla_{X}\psi=\frac{\alpha}{2}\widetilde{X}.\psi\quad\quad \forall X \in \Gamma TM
\end{eqnarray}
where $\alpha$ is a holomorphic function on $M$ and $S^{\mathbb{C}}M$ is the complex spinor bundle \cite{Rademacher}\footnote{In \cite{Rademacher} the Killing function is considered as a pure imaginary function due to the sign convention in the Clifford algebra identity $e^a.e^b+e^b.e^a=-2g^{ab}$. However our sign convention is $e^a.e^b+e^b.e^a=+2g^{ab}$ which gives a real function.}. The smooth sections of the spinor bundle admit a Hermitian symmetric inner product $(.,.)$ with adjoint involution $\mathcal{J}=\xi$ where $\xi$ is the main involutary anti-automorphism of the complex Clifford bundle which reverses the order of multiplication, so its action on a $p$-form $\omega$ is $\omega^{\xi}=(-1)^{\lfloor p/2 \rfloor}\omega$ and $j=*$ is the complex conjugation induced on the division factor of the algebra \cite{Benn and Tucker}. Keeping in mind the action of the co-derivative $\delta$ on a Clifford form $\Phi$ in the presence of torsion as $$\delta\Phi=-i_{X^a}\nabla_{X_a}\Phi+\frac{1}{2}i_{X^c}i_{X^d}T^a i_{X_c}i_{X_d}(\Phi \wedge e_a)$$ then it is easy to find the differential equations satisfied by the bilinears of the Killing spinor in the case of torsion \cite{Acik and Ertem 2015}. For $p=even$ these are given by
\begin{eqnarray}
\nabla_{X_a}(\psi\overline{\psi})_{p}&=& \alpha\, i_{X_a}(\psi\overline{\psi})_{p+1}  \\   \nonumber
d(\psi\overline{\psi})_{p}&=&\alpha (p+1)(\psi\overline{\psi})_{p+1}+T^{a}\wedge i_{X_a}(\psi\overline{\psi})_{p}   \\   \nonumber
\delta (\psi\overline{\psi})_{p}&=& -\frac{1}{2}(i_{X_c}i_{X_a}T^{b})(i_{X^a}i_{X^c}(\psi\overline{\psi})_{p})\wedge e_{b}+(i_{X_c}i_{X_a}T^{a})i_{X^c}(\psi\overline{\psi})_{p}
\end{eqnarray}
and for $p=odd$ they are
\begin{eqnarray}
\nabla_{X_a}(\psi\overline{\psi})_{p}&=& \alpha\, {e_a}\wedge(\psi\overline{\psi})_{p-1}  \\   \nonumber
d(\psi\overline{\psi})_{p}&=&T^{a}\wedge i_{X_a}(\psi\overline{\psi})_{p}   \\   \nonumber
\delta (\psi\overline{\psi})_{p}&=&-\alpha (n-p+1) (\psi\overline{\psi})_{p-1}-\frac{1}{2}(i_{X_c}i_{X_a}T^{b})(i_{X^a}i_{X^c}(\psi\overline{\psi})_{p})\wedge e_{b}-(i_{X_c}i_{X_a}T^{a})i_{X^c}(\psi\overline{\psi})_{p}.
\end{eqnarray}
We focus on the scalar and 1-form part of the inhomogeneous bilinear and if we denote the vector field corresponding to $\psi$ by $V_{\psi}$ then the equations are immediate
\begin{eqnarray}
d(\psi,\psi)&=&\alpha \widetilde{V_{\psi}}   \\   \nonumber
\delta (\psi,\psi)&=& 0
\end{eqnarray}
and
\begin{eqnarray}
d\widetilde{V_{\psi}}&=&T^{a}\wedge i_{X_a} \widetilde{V_{\psi}}   \\   \nonumber
\delta \widetilde{V_{\psi}}&=& -4\alpha (\psi,\psi)-i_{V_{\psi}}i_{X_a}T^a.
\end{eqnarray}
The scalar bilinear $(\psi,\psi)$ is real and from $\widetilde{V_{\psi}}=(\psi,e_a \psi)e^a=(e_a \psi,\psi)e^a=(\psi,e_a \psi)^* e^a=\widetilde{V_{\psi}}^*$, it is seen that the 1-form part is also real which means that $\alpha$ is real too. We would like to give torsion a spinorial origin so that we select the torsion 2-forms as $T^a=e^a \wedge \frac{d\alpha}{\alpha}$ where $\alpha$ is the (Killing) function appearing in the Killing spinor equation and it is clear from the first line of the above equations that this function also couples the scalar and the co-vector parts of the spinor bilinear. It is interesting to see that by this choice of torsion, the convenient part of the above set of equations are consistent. The consistency of the first lines can be seen from $$d\widetilde{V_{\psi}}=T^{a}\wedge i_{X_a} \widetilde{V_{\psi}}=e^a \wedge \frac{d\alpha}{\alpha}\wedge i_{X_a}\widetilde{V_{\psi}}=-\frac{d\alpha}{\alpha}\wedge \widetilde{V_{\psi}}$$ so, $\alpha\,d\widetilde{V_{\psi}}+d\alpha \wedge \widetilde{V_{\psi}}=0$ which means that $d(\alpha \widetilde{V_{\psi}})=0$ and this is in harmony with what we have in the first line of equation (33). The second line of the equation (34), $\alpha \delta (\widetilde{V_{\psi}})=-4\alpha^2 (\psi,\psi)-3V_{\psi}(\alpha)$ can further be analyzed so as to get a conserved 1-form current for our specific spinor field, i.e. the conditions that will arise by taking $\delta \widetilde{V_{\psi}}=0$. To do this we use the definition of the action of the co-derivative on $\alpha \widetilde{V_{\psi}}$ and using the form of the torsion we get
$$\delta(\alpha \widetilde{V_{\psi}})=-i_{X^a}\nabla_{X_a}(\alpha \widetilde{V_{\psi}})+\frac{1}{2}(g^{da}X^c(\alpha)-g^{ca}X^d(\alpha))i_{X_c}i_{X_d}(\widetilde{V_{\psi}}\wedge e_a)$$
and we obtain $$\delta(\alpha \widetilde{V_{\psi}})=-\alpha i_{X^a}\nabla_{X_a}\widetilde{V_{\psi}}-4i_{\widetilde{d\alpha}}\widetilde{V_{\psi}}.$$ From the first line of (32) for $p=1$ we reach $\delta(\alpha \widetilde{V_{\psi}})=-4(\alpha^2 (\psi,\psi)-i_{\widetilde{d\alpha}}\widetilde{V_{\psi}})$ and finally we have $$\delta(\alpha \widetilde{V_{\psi}})=\alpha \delta \widetilde{V_{\psi}}-g(d\alpha,\widetilde{V_{\psi}})$$.
 We also can assume a form for curvature 2-forms as ${R^{a}}_{b}=f e^a \wedge e_b$. From the known action of curvature operator on spinor fields we get $$\mathbf{R}(X_c,X_d)\psi=-\frac{1}{4}i_{X_c}i_{X_d}R_{ab}e^{ab}\psi=-\frac{f}{4}(g_{da}g_{cb}-g_{db}g_{ca})e^{ab}\psi=\frac{f}{2}e_{cd}\psi$$
and also from equation (30) we have $$\mathbf{R}(X_c,X_d)\psi=\frac{1}{2}[X_c(\alpha)e_d-X_d(\alpha)e_c].\psi+\frac{\alpha^2}{2}e_{dc}.\psi$$ and as a result we deduce a constraint equation
\begin{eqnarray}
\frac{1}{2}[X_c(\alpha)e_d-X_d(\alpha)e_c].\psi=0
\end{eqnarray}
and the equality
\begin{eqnarray}
f=-\alpha^2.
\end{eqnarray}
Equation (35) also implies that $d\alpha.\psi=0$. From what we have set up, we can prominently reduce the equations of motion of our relativistic membrane step by step and can comment on the results.\\

First it should be useful to see that with our choice of torsion the second fundamental form is still symmetric: From (B6) we have $$g(\mathbf{T}(X_i,X_j),N)=H(X_i,X_j)-H(X_j,X_i)$$ and since $\widetilde{N}=e^1$ we can write $$H_{ij}-H_{ji}=e^1(\mathbf{T}(X_i,X_j))=2T^1(X_i,X_j)=i_{X_j}i_{X_i}(e^1\wedge \frac{d\alpha}{\alpha})=0$$ hence the claim. It is also important to note that the covariant divergence of the total stress-energy-momentum tensor is seen to be exact by our set up; so $$\nabla.T=-*^{-1}(T^q\wedge i_{X_q}R_{ab}\wedge *{e^{abp}})e_p=-\alpha*^{-1}(d\alpha \wedge e_{ab}\wedge *(e^{abp}))e_p=d(-6\alpha^2).$$ Before going further we may note that the last term at the left hand side of both equations (23) and (29) is $i_{N}(\nabla.T)=-6N(\alpha^2)$. So, the equation of motion in the presence of torsion and curvature for the \textbf{Dirac bubble} coupled to an electromagnetic field is
\begin{eqnarray}
\kappa \mathcal{H}+6N(\alpha^2)/|d\Phi|=|F_{+}|^2/2.
\end{eqnarray}

To reach the equation of motion of the \"{O}nder-Tucker bubble in this non-Riemannian spacetime we have to manipulate the terms in equation (29) separately. From (B4) we can contract the Gauss equation to get $$Ric(Y,Z)=\widehat{Ric}(Y,Z)+R(N,Y,Z,N)+H(Y,X_i)H(X^i,Z)-H(Y,Z)\mathcal{H}$$
and one more contraction yields the Codazzi relation
\begin{eqnarray}
\mathcal{R}=\widehat{\mathcal{R}}+2Ric(N,N)+H_{ji}H^{ij}-\mathcal{H}^2.
\end{eqnarray}
This equation accounts for the first three terms in (29). The next term $g(\breve{\Pi}_N\mathbf{T}(X_a,X_b),\breve{\Pi}_Nh(X^b,X^a))$ reads $g(\mathbf{T}(X_j,X_i),h(X^i,X^j))$ where the indices $i$ and $j$ run through $0,2,3$. Since $h(X^i,X^j)=H^{ij}N$ we should use the normal part of $\mathbf{T}(X_i,X_j)=\nabla_{X_i}X_j-\nabla_{X_j}X_i-[X_i,X_j]$; that is $\mathbf{T}(X_i,X_j)^\perp=(H_{ij}-H_{ji})N$ which vanishes according to our choice of the torsion tensor; these follow from the two equations following (36). So, the equation of motion in the presence of torsion and curvature for the \textbf{\"{O}nder-Tucker bubble} coupled to an electromagnetic field is
\begin{eqnarray}
\kappa (\mathcal{H}^2-H^2)+6N(\alpha^2)/|d\Phi|=|F_{+}|^2/2,
\end{eqnarray}
where $H^2=H_{ij}H^{ij}$. This could also be rewritten by using the Codazzi relation (38) as
\begin{eqnarray}
\kappa (\mathcal{\widehat{R}}+6 \alpha^2)+6N(\alpha^2)/|d\Phi|=|F_{+}|^2/2.
\end{eqnarray}

 \section{Conclusion}

We analyzed the equations of motion of membranes under the action of external fields and under the influence of curvature and torsion. Our results specifically generalize the pioneering work of Dirac and also the invaluable work of \"{O}nder and Tucker for the terms coming from torsion. Although Dirac's work and also \"{O}nder and Tucker's work were based on an action functional, our treatment assumed a total stress tensor for determining the motion of membranes. We also restricted our attention to four dimensional space-times and codimension one immersions; so the possible contributions that could come from the normal bundle connection forms disappeared automatically. Selecting a spherical topology for the membrane and coupling it to an external electromagnetic field and then fixing the geometric stress tensor of the time-like hypersurface have given equations of motion with extra curvature and torsion terms in comparison to Dirac's and \"{O}nder and Tucker's work respectively.

In Dirac's paper the Coulombic stresses were balanced by the extrinsic curvature of the immersion, though in \"{O}nder and Tucker's work these stresses were balanced by the intrinsic (Gaussian) curvature of the immersion. In our work there are also torsion terms, where in the absence of electromagnetic coupling, could counteract the remaining terms and stabilize the membrane dynamics in a different manner. For Riemann-Cartan backgrounds admitting generalized Killing spinors we have used the Killing function as the source of torsion and reached equations dependent on this function. It is clear from the equations of motion that if the gradient of Killing function has no normal component with respect to the hypersurface, our result reduces to that of Dirac's; but on the contrary the reduction to \"{O}nder and Tucker's case is impeded by the remaining quadratic term of this function.

The next task should be to answer some important open questions in this setup. A Lagrangian formulation of this construction is crucial, which in the classical domain may generate the form of the hypersurface stress tensor by the usual Noetherian procedure. The motion of membranes in higher-dimensional space-times and with higher codimensions would be of interest, so that the contribution coming from the existence of the external twist potential could be analyzed and interpreted seperately. Other interesting topologies could be selected for the membrane, for example the ones supporting axially symmetric configurations and also the membrane could be coupled to a gravitational field so that the background becomes active. For such a gravitating membrane the distributional Einstein-Cartan equations come into play. Another problem should be to deduce the spin structure of the membrane's history and investigate the properties of spinor fields on the hypersurface induced from the generalized Killing spinor. These spinor fields may generate energy minimising calibrations for the hypersurface.

\appendix

\section{Curved space-time distributions}
All the details of this appendix can be found in the reference \cite{Hartley et.al.}. Let $M$ be an n-dimensional paracompact manifold without a metric structure then the test functions on $M$ are smooth functions with compact support; also the real or complex valued continuous linear functionals over the space of test functions are called the scalar distributions on $M$. The effect of the scalar distribution $D$ on the test function $p$ is denoted as $D:p\rightarrow D\lfloor p\rfloor$. Similarly tensors on $M$ with compact support are called test tensors and the real or complex valued linear functionals over the space of these quantities are called tensor distributions with effect $T: U\rightarrow T\lfloor U\rfloor$; where the tensorial type of $T$ is dual to that of the tensor $U$. If $f$ is a function, $X$ a vector field, $S$ a tensor and $W$ is another tensor that has dual type with respect to $S$ and $T$ is a tensor distribution then the following equalities are well defined:
\begin{eqnarray}
(fT)\lfloor U\rfloor&=&T\lfloor fU\rfloor,  \\ \nonumber
(S \otimes T)(W \otimes U)&=&T\lfloor \langle S,W\rangle U\rfloor, \\ \nonumber
T_{X}\lfloor U\rfloor&=&T\lfloor X\otimes U\rfloor. \nonumber
\end{eqnarray}
Generally the tensor products of tensor distributions are not well defined. If the local frame $\{X_a\}$ and the dual co-frame $\{e^a\}$ are defined in an open subset $\mathcal{N}$ of $M$ then the components of $T$ are the scalar distributions defined in an obvious manner. If $\alpha$ is a co-vector distribution then its component distributions are given by
$$\alpha_{a}\lfloor p\rfloor=\alpha\lfloor pX_a\rfloor$$
whereas $supp(p) \subset \mathrm{N}$. For example on a test vector $V$ with $supp(V) \subset \mathcal{N}$ the $\alpha \lfloor V\rfloor=\alpha_{a} e^a\lfloor V\rfloor$ equality holds. The components transform in the usual way under frame changes so tensor distributions can equally be regarded as distribution-valued tensors. As an additional structure if there is a globally defined n-form (volume form) $\omega$ on $M$ then a distribution $\widehat{S}$ can be defined by
\begin{eqnarray}
\widehat{S}\lfloor U\rfloor=\int_{M} \langle S,U\rangle \omega. \nonumber
\end{eqnarray}
Let $M$ be orientable and be divided into two disjoint open regions $M^+$ and $M^-$ by an (n-1)-dimensional submanifold $\Sigma$, defined locally by the equation $\Phi=0$ with $d\Phi\neq0$ in a neighborhood of $\Sigma$. The orientation is fixed by $\Sigma=\partial M^-=-\partial M^+$. While the Heaviside scalar distribution is defined as
\begin{eqnarray}
\Theta^{\pm}\lfloor p\rfloor=\int_{M^\pm}p \omega
\end{eqnarray}
the Dirac 1-form distribution is defined by
\begin{eqnarray}
\delta\lfloor V\rfloor=\int_{\Sigma} i_{V}\omega.
\end{eqnarray}
The Leray form $\sigma$ is defined by the equality $\omega=d\Phi \wedge \sigma$ in a neighborhood of $\Sigma$ and this makes it possible to define the usual scalar Dirac distribution as
\begin{eqnarray}
\delta(\Phi)\lfloor p\rfloor=\int_{\Sigma}p \sigma.
\end{eqnarray}
One can show that $\delta=\delta(\Phi)d\Phi$ and $d\Theta^{\pm}=\pm \delta$.

A function $f$ on $M$ is regularly $C^k$ discontinuous at $\Sigma$ if $f$ and its first k derivatives are continuous on $M^{\pm}$ and converge uniformly to limits $f^{\pm}_{\Sigma}$ etc. at $\Sigma$. A regularly discontinuous tensor $S$ is one whose components in any given chart intersecting $\Sigma$ are regularly discontinuous functions. The discontinuity $[S]$ of $S$ is an ordinary continuous tensor over $\Sigma \subset M$ defined by
\begin{eqnarray}
[S]=S^+_{\Sigma}-S^-_{\Sigma}.
\end{eqnarray}
Let $\widehat{f}$ be the distribution related to a regularly discontinuous function $f$. If $f^{\pm}$ are arbitrary smooth extensions of $f|_{M^{\pm}}$ to $M$ then
\begin{eqnarray}
\widehat{f}=\Theta^{+}f^{+}+\Theta^{-}f^{-}.
\end{eqnarray}
It follows that
\begin{eqnarray}
d\widehat{f}=\Theta^{+}df^{+}+\Theta^{-}df^{-}+[f]\delta,
\end{eqnarray}
where $[f]$ is continuous on $supp(\delta)=\Sigma$. For a smooth $\nabla$ and regularly discontinuous tensor $S$ we have
\begin{eqnarray}
\nabla \widehat{S}=\Theta^{+}\nabla S^{+}+\Theta^{-}\nabla S^{-}+\delta\otimes [S],
\end{eqnarray}
and if both $\nabla$ and $S$ are regularly discontinuous then
\begin{eqnarray}
\nabla \widehat{S}=\Theta^{+}\nabla^+ S^{+}+\Theta^{-}\nabla^- S^{-}+\delta\otimes [S].
\end{eqnarray}

\section{Gauss-Codazzi Formalism with torsion}
We here give the embedding equations for a hypersurface into a space-time with torsion. If $(M,g,\nabla)$ is the ambient space-time then the structure of the regular hypersurface is given by the triple $(\Sigma, \widehat{g}, \widehat{\nabla})$ where $\widehat{g}=\breve{\Pi}_N g$ is the induced metric (the first fundamental form) which is compatible with the induced connection $\widehat{\nabla}$. The Gauss formula yields for a pair of tangent vector fields $X,Y$ to $\Sigma$; namely $X,Y \in \Gamma T\Sigma$
\begin{eqnarray}
\nabla_{X}Y=\widehat{\nabla}_{X}Y+h(X,Y)
\end{eqnarray}
where $h=HN$ is the shape operator of the immersion, $H$ its second fundamental form and $N$ is the unit normal field. The Weingarten formula generally is given by
\begin{eqnarray}
\nabla_{X}N=-A_{N}X+\nabla^{\bot}_{X}N
\end{eqnarray}
where $A_{N}$ is the Weingarten map satisfying $g(h(X,Y),N)=g(A_{N}X,Y)=H(X,Y)$ and $\nabla^{\bot}$ is the induced connection on the normal bundle. We will see that in the presence of torsion the Weingarten map will loose its symmetry as an endomorphism on each tangent space for the hypersurface. For the special case where $N$ has constant norm then the term $\nabla^{\bot}_{X}N$ vanishes automatically. So the Weingarten formula reads
\begin{eqnarray}
\nabla_{X}N=-A_{N}X.
\end{eqnarray}
If $\mathbf{R}$ is the curvature operator and $\mathbf{T}$ torsion operator of $\nabla$, $\mathbf{\widehat{R}}$ the curvature operator and $\mathbf{\widehat{T}}$ the torsion operator of $\widehat{\nabla}$ then one can easily find the below equations
\begin{eqnarray}
(\mathbf{R}(X,Y)Z)^{\|}=\mathbf{\widehat{R}}(X,Y)Z+H(X,Z)A_{N}Y-H(Y,Z)A_{N}X \nonumber
\end{eqnarray}
and
\begin{eqnarray}
(\mathbf{R}(X,Y)Z)^{\bot}=\{(\nabla_{X}H)(Y,Z)-(\nabla_{Y}H)(X,Z)+H(\mathbf{\widehat{T}}(X,Y),Z)\}N+H(Y,Z)\nabla^{\bot}_{X}N-H(X,Z)\nabla^{\bot}_{Y}N \nonumber
\end{eqnarray}
for non-constant $g(N,N)$ and also from
\begin{eqnarray}
\mathbf{T}(X,Y)=\mathbf{\widehat{T}}(X,Y)+h(X,Y)-h(Y,X) \nonumber
\end{eqnarray}
we reach
\begin{eqnarray}
(\mathbf{T}(X,Y))^{\|}=\mathbf{\widehat{T}}(X,Y) \nonumber
\end{eqnarray}
and
\begin{eqnarray}
(\mathbf{T}(X,Y))^{\bot}=h(X,Y)-h(Y,X). \nonumber
\end{eqnarray}
So the Gauss equation is
\begin{eqnarray}
R(X,Y,Z,W)=\widehat{R}(X,Y,Z,W)+g(h(Y,W),h(X,Z))-g(h(X,W),h(Y,Z))
\end{eqnarray}
that is left unchanged in the presence of torsion where $R (\widehat{R})$ is the Riemann tensor of $M (\Sigma)$. The Codazzi equation becomes
(with unit $N$)
\begin{eqnarray}
R(X,Y,Z,N)=(\nabla_{X}H)(Y,Z)-(\nabla_{Y}H)(X,Z)+H(\mathbf{\widehat{T}}(X,Y),Z)
\end{eqnarray}
and also we have
\begin{eqnarray}
g(\mathbf{T}(X,Y),N)N=h(X,Y)-h(Y,X).
\end{eqnarray}

\acknowledgments
This work is supported by the Scientific and Technological Research Council of Turkey (T\"{U}BİTAK) Research Project No. 118F086.

 \end{document}